\documentclass[12pt]{article}
\usepackage{graphics,epsfig,a4wide,amsmath,rotating}
\newcommand{\Ld}{\Lambda}
\newcommand{\Gm}{\Gamma}
\newcommand{\Sg}{\Sigma}
\newcommand{\Sgs}{\Sigma^*}
\newcommand{\Xs}{\Xi^*}
\newcommand{\eps}{\epsilon}
\newcommand{\bk}{\bar{K}}

\newcommand{\be}{\begin{equation}}
\newcommand{\ee}{\end{equation}}
\newcommand{\ba}{\begin{eqnarray}}
\newcommand{\ea}{\end{eqnarray}}
\newcommand{\rw}{\rightarrow}
\newcommand{\pr}{^{\,\prime}}

\newcommand{\del}{\partial}
\newcommand{\rgl}{\rangle}
\newcommand{\tp}{{\tilde p}}
\newcommand{\tph}{{\tilde \phi}}
\begin{document}

\title{Chiral coupled channel dynamics of the $\Ld(1520)$ and the $K^-p\rw\pi^0\pi^0\Ld$
reaction}

\author{Sourav Sarkar, E. Oset and M.J. Vicente Vacas\\
{\small Departamento de F\'{\i}sica Te\'orica and IFIC,
Centro Mixto Universidad de Valencia-CSIC,} \\
{\small Institutos de
Investigaci\'on de Paterna, Aptd. 22085, 46071 Valencia, Spain}\\
}
\maketitle
\begin{abstract}
We study the $\Lambda(1520)D_{03}$ in a chiral coupled channel approach.
This resonance appears as dynamically generated from the interaction of
the decuplet of baryons and the octet of mesons in $s$-wave, and its treatment is improved here
with the phenomenological inclusion
of the $\bar{K} N$ and $\pi \Sigma$ channels in $d$-wave. Since 
the most important building block in the 
$\Lambda(1520)$ is the $\pi \Sgs(1385)P_{13}$ channel, we study the $K^- p \to \pi
\Sgs(1385) (\pi^0 \Lambda)$ reaction in the region of the $\Lambda(1520)$ and
above, and compare the results with recent experimental data.  With the coupling 
of the $\Lambda(1520)$ to the $\pi \Sgs$ channel predicted by the theory
we find a cross section in good agreement with the data and there is as well agreement
for the invariant mass distributions which show a neat peak for the  
$\Sgs(1385)$ in the $(\pi^0 \Lambda)$ spectrum.  Predictions are made of a
strong $\Lambda(1520)$ resonant peak of the cross section, as a function of the 
$K^-$ momentum, in the region below the measured data which, if confirmed
experimentally, would give a stronger support to the idea of the $\Lambda(1520)$
as a dynamically generated resonance.
\end{abstract}
\newpage

\section{Introduction}

The chiral coupled channel approach, implementing exact unitarity in coupled channels
and using input from chiral Lagrangians, has allowed to make predictions beyond the 
restricted range of energies of chiral perturbation theory and is having a
great impact in the study of meson baryon interaction at low energies.
At the same time it has shown that many known resonances listed by the Particle Data 
Group (PDG)~\cite{Eidelman:2004wy} 
 qualify as dynamically generated, or in simpler words,
they  are quasibound states of a meson and a baryon. After early studies in this
direction showing that the $\Lambda (1405)S_{01}$ and the $N^*(1535)S_{11}$ were dynamically
generated resonances~\cite{Kaiser:1995cy,Kaiser:1996js,angels,Nacher:1999vg,oller,Inoue:2001ip,
bennhold,Garcia-Recio:2002td},
 more
systematic studies have shown that there are two octets and one singlet of
resonances from the interaction of the octet of pseudoscalar mesons with the
octet of stable baryons~\cite{Jido:2003cb,Garcia-Recio:2003ks}. 
   
   Further work in this direction~\cite{lutz,decu_ss} has
shown that many of the $3/2^-$ low lying baryonic resonances appear as
dynamically generated from the interaction of the decuplet of baryons and the
octet of mesons. Clear peaks in the amplitudes and poles in the complex plane appear 
for states that can be associated to the $\Delta(1700)D_{33}$, $\Sigma(1670)D_{13}$, 
$\Sigma(1940)D_{13}$ and $\Xi(1820)D_{13}$, while the $N^*(1520)D_{13}$ and $\Lambda(1520)D_{03}$
are reproduced only qualitatively hinting at the relevance of extra coupled channels.
  In particular the $\Lambda(1520)$ appears displaced in mass, around 1560 MeV in
 \cite{lutz} and 1570 MeV in~\cite{decu_ss}.
In the chiral coupled channel approach of Refs.~\cite{lutz,decu_ss}
 this resonance couples to the $\pi \Sgs(1385)$ and 
$K \Xi^*(1530)P_{13}$ channels, particularly to
the former one. With the $\pi^+ \Sg^{*-}$, $\pi^- \Sg^{*+}$, $\pi^0 \Sg^{*0}$ masses
7 MeV above, 2 MeV above and 1 MeV below the nominal $\Lambda(1520)$ mass and the
strong coupling of the resonance to $\pi \Sgs$, the state could qualify as a loosely 
bound $\pi \Sgs$ state.
  However, the lack of other relevant channels which couple
to the quantum numbers of the resonance makes the treatment 
of~\cite{lutz,decu_ss} only semiquantitative. In
particular, the $\Lambda(1520)$ appears in~\cite{lutz,decu_ss}
at  higher energy than the nominal one and with a  large width of
about 130 MeV, nearly ten times larger than the physical width.
  This large width is a necessary consequence of the large coupling to the 
$\pi \Sgs$ channel and the fact that the pole appears at energies above
the $\pi \Sgs$ threshold. On the other hand, if we modify the
subtraction constants of the meson baryon loop function to bring the pole below
the $\pi \Sgs$ threshold, then the pole appears without imaginary part.
Since the width of the $\Lambda(1520)$ resonance comes basically from the
decay into
the $\bar{K} N$ and $\pi \Sigma(1193)$, the introduction of these channels is 
mandatory to reproduce the shape of the $\Lambda(1520)$ resonance.

   In the present work we include the $\bar{K} N$ and $\pi \Sigma$ channels into
the set of coupled channels which build up the $\Lambda(1520)$.
This is done phenomenologically with no links with chiral Lagrangians.
 The novelty with
respect to the other channels already accounted for~\cite{lutz,decu_ss},
 which couple in $s$-wave,
 is that the new channels couple in $d$-waves. Fitting two parameters to the partial decay widths of the  $\Lambda(1520)$
into $\bar{K} N$ and $\pi \Sigma$, a good shape for the $\Lambda(1520)$ 
dominated 
amplitudes is obtained at the right position and with the proper experimental
width.  The coupling of the $\Lambda(1520)$ to the $\pi \Sgs$ 
channel is a prediction of the theory and we use this to study
the reaction $K^- p \to \pi \Sigma(1385) (\pi^0 \Lambda)$
which is closely related to the strength of 
this coupling. We then compare with recent
experimental results measured above the $\Lambda(1520)$ energy. The
agreement with the data is good and the cross section is sizeable thanks to
the large coupling of the $\Lambda(1520)$ to the $\pi \Sgs$ channel. Other
 standard mechanisms for the  
$K^- p \to \pi^0  \pi^0 \Lambda$ reaction without the $\Lambda(1520)$
give too small a cross section compared
to experiment in a wide range of energies around the $\Lambda(1520)$ peak.  

We also compare the invariant mass distributions for  
$\pi^0 \Lambda$, where a distinct peak associated to the $\Sigma(1385)$ resonance
is seen, in good agreement with experiment. 

   We also make predictions for the cross section for $K^- p$ energies around
the $\Lambda(1520)$, where we find a large peak with the 
$\Lambda(1520)$ shape, not measured so far, and which, if confirmed 
experimentally, would
give a strong support to the idea of the $\Lambda(1520)$ as a dynamically
generated resonance.

   Inclusion of the new channels $\bar{K} N$ and $\pi \Sigma$ into
the coupled channel approach allows one to calculate
the cross sections of other reactions
where the  $\Lambda(1520)$ appears, much as has been the case of the
$\Lambda(1405)$~\cite{review}, where
a large variety of reactions could be studied within the chiral unitary approach
taking into account that the $\Lambda(1405)$  is dynamically generated from the 
$\bar{K}N$ and coupled channels in $s$-wave.

\section{Decuplet octet interaction and the $\Ld(1520)$}

Following~\cite{decu_ss}, we briefly recall how the $\Ld(1520)$ is generated 
dynamically in the $s$-wave interaction of the decuplet of baryons with the octet 
of pseudoscalar mesons. We consider the lowest order term of the 
chiral Lagrangian given by~\cite{manohar}
\be
{\cal L}=-i\bar T^\mu {\cal D}\!\!\!\!/ T_\mu 
\label{lag1} 
\ee
where $T^\mu_{abc}$ is the decuplet of Rarita Schwinger fields and ${\cal D}^{\nu}$ the covariant derivative
given by
\be
{\cal D}^\nu T^\mu_{abc}=\del^\nu T^\mu_{abc}+(\Gamma^\nu)^d_aT^\mu_{dbc}
+(\Gamma^\nu)^d_bT^\mu_{adc}+(\Gamma^\nu)^d_cT^\mu_{abd}
\ee
with $\mu$ the Lorentz index and $a,b,c$ the $SU(3)$ indices.
The vector current $\Gm^\nu$ is given by
\be
\Gm^\nu=\frac{1}{2}(\xi\del^\nu \xi^\dagger+\xi^\dagger\del^\nu \xi)
\ee
with
\be
\xi^2=U=e^{i\sqrt{2}\Phi/f}
\ee 
where $\Phi$ is the 3$\times$3 matrix of fields for the pseudoscalar 
mesons~\cite{Gasser:1984gg} and $f=93$ MeV. Consideration of only the $s$-wave part
of the baryon meson interaction and the use of non-relativistic approximations
as described in detail in~\cite{decu_ss} allows for substantial technical
simplifications, and writing  $T_\mu  \equiv T u_\mu$, with $u_\mu$ the 
Rarita Schwinger spinor, the Lagrangian can be written as the flavor trace
\be
{\cal L}=3i\, Tr \left[  \bar{T}\cdot T\, \Gamma^{0T }\right]
\label{lag1xx} 
\ee
where
\be
 \left(  \bar{T}\cdot T\right)^{a}_d=\sum_{b,c}  \bar{T}^{abc}T_{dbc}
\ee
and $\Gamma^{0T }$ is the transposed matrix of $\Gamma^{0 }$ with 
$\Gamma^{\nu}$ given, up to two meson fields, by
\be
\Gamma^{\nu} =\frac{1}{4 f^2}\left( \Phi\partial^{\nu}\Phi-\partial^{\nu}\Phi\Phi\right).
\ee
 From the Lagrangian of Eq. (\ref{lag1xx}) and with the ordinary correspondence of the
${T}^{abc}$ components to the decuplet fields used in \cite{decu_ss}
we obtain the $s$-wave transition amplitudes for a meson of 
incoming and outgoing momentum $k$ and $k\pr$ respectively as
\be
V_{ij}=-\frac{1}{4f^2}C_{ij}(k^0+k^{\pr 0}).
\label{poten}
\ee
For the quantum numbers $S=-1$ and $I=0$ the relevant channels are
$\pi\Sgs$ and $K\Xs$. The corresponding coefficients $C_{ij}$
are shown in table~\ref{tabS-1I0} where we have used the isospin
states\footnote{we use $|\pi^+\rgl=-|1\ 1\rgl$}
\ba
|\pi\Sgs ;\ I=0\rgl&=&\frac{1}{\sqrt{3}}\ |\pi^-\Sigma^{*+}\rgl
-\frac{1}{\sqrt{3}}\ |\pi^0\Sigma^{*0}\rgl
-\frac{1}{\sqrt{3}}\ |\pi^+\Sigma^{*-}\rgl\nonumber\\
|K\Xs ;\ I=0\rgl&=&-\frac{1}{\sqrt{2}}\ |K^0\Xi^{*0}\rgl
+\frac{1}{\sqrt{2}}\ |K^+\Xi^{*-}\rgl~.
\ea
 
\begin{table}[h]
\begin{center}
\begin{tabular}{c|cc}
\hline
 & $\pi\Sgs$ & $K\Xs$ \\	
\hline 
$\pi\Sgs$ & 4 & $-\sqrt{6}$ \\
$K\Xs$ & $-\sqrt{6}$ & 3 \\
\hline
\end{tabular}
\caption{$C_{ij}$ coefficients for $S=-1$, $I=0$.}
\label{tabS-1I0}
\end{center}
\end{table}

The matrix $V$ is then used as the kernel of the Bethe-Salpeter equation to
obtain the unitary transition matrix~\cite{angels}. This results in the matrix equation
\be
T=(1-VG)^{-1}V
\label{BS}
\ee
where $G$ is a diagonal matrix representing the meson-baryon loop function
\begin{eqnarray}
G_{l}&=& i \, 2 M_l \int \frac{d^4 q}{(2 \pi)^4} \,
\frac{1}{(P-q)^2 - M_l^2 + i \epsilon} \, \frac{1}{q^2 - m^2_l + i
\epsilon}  \nonumber \\ 
&=& \frac{2 M_l}{16 \pi^2} \left\{ a_l(\mu) + \ln
\frac{M_l^2}{\mu^2} + \frac{m_l^2-M_l^2 + s}{2s} \ln \frac{m_l^2}{M_l^2}
-2i\pi \frac{q_l}{\sqrt{s}}
\right. \nonumber \\ & &  \phantom{\frac{2 M}{16 \pi^2}} +
\frac{q_l}{\sqrt{s}}
\left[
\ln(s-(M_l^2-m_l^2)+2 q_l\sqrt{s})+
\ln(s+(M_l^2-m_l^2)+2 q_l\sqrt{s}) \right. \nonumber  \\
& & \left. \phantom{\frac{2 M}{16 \pi^2} +
\frac{q_l}{\sqrt{s}}}
\left. \hspace*{-0.3cm}- \ln(s-(M_l^2-m_l^2)-2 q_l\sqrt{s})-
\ln(s+(M_l^2-m_l^2)-2 q_l\sqrt{s}) \right]
\right\},
\label{propdr}
\end{eqnarray}
in which $M_l$ and $m_l$ are the masses of the baryons and mesons respectively,
$s=P^2$, with $P$ the total four momentum of the meson baryon system and $q_l$ denotes the 
three-momentum of the meson or baryon in the center of mass frame.
In the second equality we have removed an infinity, that we obtain
for instance evaluating the integral with dimensional regularization.  In getting 
the finite expression at a  regularization scale $\mu$, we are implicitly assuming that there is
a higher order counterterm that has canceled the infinity and provided a remnant finite part
which is the subtraction constant $a_l(\mu) $. In as much as the $a_l(\mu) $ will be a fit parameter
in the theory there is no need to use explicitly the counterterm Lagrangians. Alternatively one can 
think of a cut off regularization without using higher order terms. A cut off in the three momentum of 
the order of 1 GeV is what we would call natural size in this case, and then it was  proved in
\cite{oller} that this regularization procedure was equivalent to that of Eq.~(\ref{propdr})
using $\mu\approx 700$ MeV and $a_l(\mu)\approx -2$.
One then looks for poles of the transition matrix $T$ in the complex $\sqrt{s}$ 
plane. The complex poles, $z_R$, appear in unphysical Riemann sheets.
A good relationship between the real and twice the imaginary part of the complex pole positions
with the mass and width of the associated Breit Wigner shapes in the real axis is obtained
in what we call the second Riemann sheet. This is defined by taking $G_l$ from Eq.~(\ref{propdr})
(which is the expression for $G$ for the first Riemann sheet) and substituting
\be
G_l^{2nd}=G_l+2i\,\frac{q_l}{\sqrt{s}}\,\frac{M_l}{4\pi},
\label{defR2}
\ee
where the variables on the right hand side of the above equation are
evaluated in the first (physical) Riemann sheet, for the channels which are above threshold
at an energy equal to $\rm{Re}(z)$. This prescription is equivalent to changing the sign of $q_l$
in Eq.~(\ref{propdr}) for those channels.
\begin{figure}[h]
\includegraphics[width=0.5\textwidth]{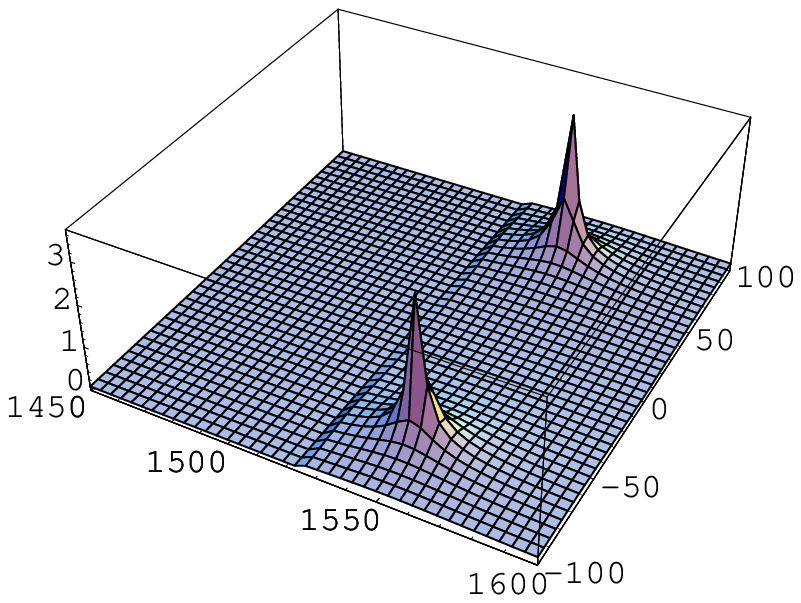}
\includegraphics[width=0.5\textwidth]{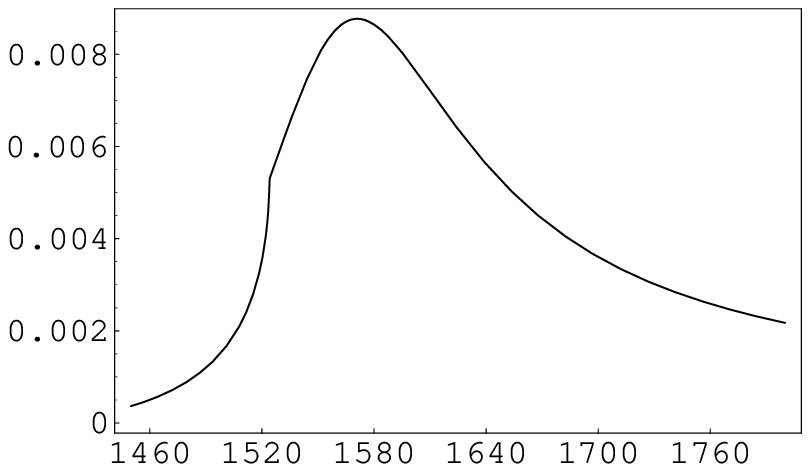}
\caption{(Color online) Left: The $\Ld(1520)$ pole as seen in the $\pi\Sgs\rw\pi\Sgs$ 
amplitude in the complex $\sqrt{s}$ plane. Right:$|T_{\pi\Sgs\rw\pi\Sgs}|^2$ in
(MeV$^{-2}$).}
\label{polefig1}
\end{figure}

Using the natural size values~\cite{oller}
$a=-2$ and $\mu=700$ MeV, we find
a pole at $z_R=1550-i67$ as seen in fig.~\ref{polefig1},
which we can well associate with the 4-star resonance
$\Ld(1520)$. The residue at this pole
indicates a strong coupling to the $\pi\Sgs$ channel~\cite{decu_ss}.
However, the experimental mass and width are lower and there are
also large branching ratios 
of the $\Ld(1520)$ to the $\bar KN$ and
$\pi\Sigma$ channels. 
In the following section we will phenomenologically
add these channels to our coupled channel scheme.
   
\section{Introduction of the $\bk N$ and $\pi\Sg$ channels}

We will generate the resonance $\Ld(1520)$ in coupled channels involving
the $\pi\Sgs$, $K\Xi^*$, $\bk N$ and $\pi\Sg$. 
However, 
we shall only couple the
$\bk N$ and $\pi\Sg$ channels to the dominant $\pi\Sgs$ channel as described below.
The lowest partial wave in which $\bar{K}N$ and $\pi\Sg$ can couple to spin parity
$3/2^-$ is $L=2$  and thus we consider these states in $d$-wave. From the point of 
view of strangeness and isospin other channels like $\eta \Lambda$ and $K \Xi$  
would be allowed (and they are considered in the $s$-wave study of $\bk N$ and
coupled channels in \cite{angels,oller,bennhold,Garcia-Recio:2002td,Jido:2003cb,Garcia-Recio:2003ks}.
However, their thresholds are at 1663 MeV and 1880 MeV respectively, such that their influence in the
region around 1520 MeV should be small. In any case, since the $\Lambda(1520)$ does not decay in these
channels their influence could only be in the mass of the resonance, not in its width, but the mass 
will be obtained by fine tunning the subtraction constant of the dominant $\pi\Sgs$ channel.

\begin{figure}[h]
\centerline{
\includegraphics[width=0.4\textwidth]{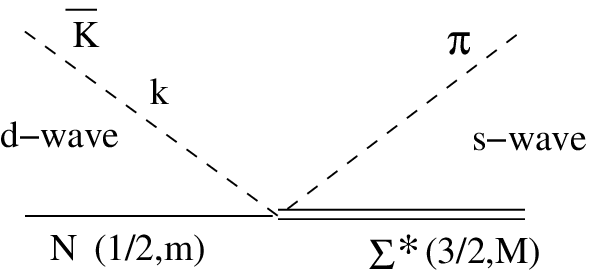}
}
\caption{The $\bk N\rw \pi\Sgs$ vertex}
\label{point}
\end{figure}

Consider the transition $\bk N$ ($d$-wave) to  $\pi\Sgs$ ($s$-wave) as shown in
fig.~\ref{point}. We start with an amplitude of the form
\be
-it_{\bk N\rw\pi\Sgs}=-i\beta_{\bk N}\ |\vec k|^2 \left[T^{(2)\dagger}
\otimes Y_{2}(\hat{k})
\right]_{0\,0}
\ee
where $ T^{(2)\, \dagger}$ is a (rank 2) spin transition operator 
defined by
\[\langle 3/2\ M|\ T^{(2)\dagger}_\mu\ |1/2\ m\rangle
={\cal C}(1/2\ 2\
3/2;m \ \mu \ M)\ \langle 3/2||\ T^{(2) \dagger}\ ||1/2\rangle~,\]
$Y_2(\hat{k})$ is the spherical harmonic coupled to $ T^{(2) \dagger}$ to
produce a scalar, and
$\vec k$ is the momentum of the $\bk$. The 3rd component of spin 
of the initial nucleon and the final $\Sgs$ are denoted by $m$
and $M$ respectively. Choosing appropriately the reduced matrix element
we obtain 
\be
-it_{\bk N\rw\pi\Sgs}=-i\beta_{\bk N}\ |\vec k|^2\ {\cal C}(1/2\ 2\
3/2;m,M-m)Y_{2,m-M}(\hat{k})(-1)^{M-m}\sqrt{4\pi}.
\ee
In the same way we
write the amplitude for $\pi\Sg$ ($d$-wave) to  $\pi\Sgs$ ($s$-wave) as
\be
-it_{\pi\Sg\rw\pi\Sgs}=-i\beta_{\pi\Sg}\ |\vec k|^2\ {\cal C}(1/2\ 2\
3/2;m,M-m)Y_{2,m-M}(\hat{k})(-1)^{M-m}\sqrt{4\pi}.
\ee

\begin{figure}
\centerline{
\includegraphics[width=0.4\textwidth]{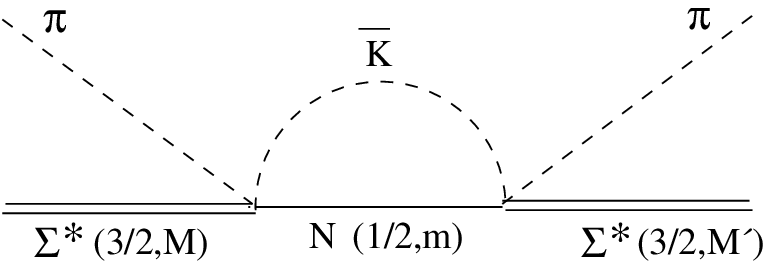}
}
\caption{$\pi\Sgs\rw\pi\Sgs$ through $\bk N$ loop arising in the Bethe-Salpeter
series}
\label{loop}
\end{figure}
Now, let us consider fig.~\ref{loop}. The loop function $G$ involving
the $\bk$ and $N$ is given by
\ba
G&=&i\int\frac{d^4q}{(2\pi)^4}\ G_N \ D_{\bk} \ 4\pi\nonumber\\
&&\beta_{\bk N}|\vec{q}|^2\sum_m{\cal C}(1/2\ 2\
3/2;m,M\pr-m)Y_{2,m-M\pr}(\hat{q})(-1)^{M\pr-m}\nonumber\\
&& \beta_{\bk N}|\vec{q}|^2\ {\cal C}(1/2\ 2\
3/2;m,M-m)Y^*_{2,m-M}(\hat{q})(-1)^{M-m}
\label{gkn}
\ea
where $G_N$ and $D_{\bk}$ are the propagators for the nucleon and the $\bk$
respectively. Eq.~(\ref{gkn}) can be further simplified by performing the angular
integration of the two spherical harmonics, which gives $\delta_{MM\pr}$ and
then using the orthogonality of the Clebsch Gordan (CG) coefficients. We obtain
\ba
G&=&i\ \delta_{MM\pr}2 M_N \int\frac{dq^0}{2\pi}\ \int\frac{|\vec q|^2\ d|\vec q|}
{(2\pi)^3}\ 4\pi\ 
(\beta_{\bk N}|\vec{q}|^2)^2\nonumber 
\frac{1}{q^2-m_K^2+i\eps}\
 \frac{1}{(P-q)^2-M_N^2+i\eps}\nonumber\\
&=&
i\ \delta_{MM\pr}2 M_N \int \frac{d^4 q}{(2 \pi)^4}\ (\beta_{\bk N}|\vec{q}|^2)^2 \ 
\frac{1}{q^2 - m^2_K + i \epsilon}\,
\frac{1}{(P-q)^2-M_N^2 + i \epsilon}~. 
\label{gkn2}
\ea
A further simplification can be done in Eq. (\ref{gkn2}) by factorizing the vertex,
$ \beta_{\bk N}|\vec{q}|^2$, on shell. This is done in the Bethe Salpeter 
approach of Ref. \cite{angels} and justified there for $s$-waves, but one finds
a more general  justification in the $N/D$ method as used in \cite{oller,nsd} which we sketch
below. Unitarity states that, above threshold,
\be
\left[ \textrm{Im}\, t^{-1}(s)\right]_{\alpha\beta}=-\frac{q_\alpha M_\alpha}{4\pi\sqrt{s}}\delta_{\alpha\beta}
\ee
Since the right hand side is $ -\textrm{Im}\, G $ one can perform a subtracted dispersion relation
and one would have 
\be
t^{-1}(s)=G(s)+V^{-1}(s)
\label{w1w2}
\ee
where $G(s)$ contains an arbitrary subtraction constant (like $a$) and $ V^{-1}(s)$ 
accounts for contact terms which remain at tree level when we remove the loops by
taking $G=0$. Eq.~(\ref{w1w2}) can be cast as 
\be
t(s)=\left[1-V(s)G(s)\right]^{-1} V(s) \;\Rightarrow \; t(s)= V(s)+ V(s) G(s)t(s)
\ee
where the last equation is the Bethe Salpeter equation except  that $V(s) t(s)$ factorize
outside the loop integral of the $VGT$ term. The caveat in the derivation is that we have only 
included the right hand cut in the dispersion relation. In as much as the contribution of the left hand 
cut
is negligible, which is the case in the meson baryon interaction since the energies of this cut
are very far from those in the real channel \cite{oller}, the on shell factorization is justified.
In fact the caveat is less restrictive because it is sufficient that the energy dependence of the 
left cut contribution is negligible in the region of interest to justify the on shell prescription,
and the contribution of the left hand cut can be absorbed into the subtraction constants.

Factorizing the vertex, i.e. $|\vec{q}|^2$, on shell
results in the
simplification that we can use the transition matrix elements
\ba
V_{\bk N\rw\pi\Sgs}&=&\beta_{\bk N}|\vec{q}_{on}|^2\nonumber\\
V_{\pi\Sg\rw\pi\Sgs}&=&\beta_{\pi\Sg}|\vec{q}_{on}\pr|^2  
\ea
where $\vec{q_{on}}$ and 
$\vec{q}_{on}\pr$ are the (on-shell) CM momenta of the $\bk$ and $\pi$ 
respectively for a given value of $s$. After removing the factor $(\beta_{\bk N}|\vec{q}|^2)^2$ in
Eq.~(\ref{gkn2}), the rest of the formula is the ordinary $G$ function for the
$s$-wave meson baryon interaction, Eq.~(\ref{propdr}). This allows us to use the same
formalism as in ordinary $s$-wave scattering assuming an effective transition
potential $\beta_{\bk N}|\vec{q_{on}}|^2$ for $\pi\Sgs\rw\bk N$.

With the matrix $V$ now given by
\be
V=\left| 
\begin{array}{cccc}
V_{\pi\Sgs\rw\pi\Sgs} & V_{\pi\Sgs\rw K\Xs} & \beta_{\bk
N}|\vec{q}_{on}|^2 & \beta_{\pi\Sg}|\vec{q}\pr_{on}|^2 \\
V_{K\Xs\rw\pi\Sgs} & V_{K\Xs\rw K\Xs} & 0 & 0 \\
\beta_{\bk N}|\vec{q}_{on}|^2 & 0 & 0 & 0 \\
\beta_{\pi\Sg}|\vec{q}\pr_{on}|^2  & 0 & 0 & 0 
\end{array}
\right|~,
\ee
we solve Eq.~(\ref{BS}) to obtain
the amplitudes $T$.
The actual transition amplitudes are related to $T$ through the
following relations
\ba
t_{\pi\Sgs\rw\pi\Sgs}&=&T_{\pi\Sgs\rw\pi\Sgs}\nonumber\\
t_{\bk N\rw\pi\Sgs}&=&T_{\bk N\rw\pi\Sgs}\ {\cal C}(1/2\ 2\
3/2;m,M-m)Y_{2,m-M}(\hat{k})(-1)^{M-m}\sqrt{4\pi}\nonumber\\
t_{\pi\Sg\rw\pi\Sgs}&=&T_{\pi\Sg\rw\pi\Sgs}\ {\cal C}(1/2\ 2\
3/2;m,M-m)Y_{2,m-M}(\hat{k})(-1)^{M-m}\sqrt{4\pi}\nonumber\\
t_{\bk N\rw\bk N}&=&T_{\bk N\rw\bk N}\ \sum_M{\cal C}(1/2\ 2\
3/2;m,M-m)Y_{2,m-M}(\hat{k})\nonumber\\
&& {\cal C}(1/2\ 2\
3/2;m\pr,M-m\pr)Y_{2,m\pr-M}^*(\hat{k}\pr)(-1)^{m\pr-m}\ 4\pi~.
\label{actual}
\ea

We then look for poles in the 2nd Riemann sheet of the complex plane. Assuming 
that the pole corresponding to the $\Ld(1520)$ appears at $z=z_R$ where $z$ stands
for the (complex) CM energy, the amplitudes close to the pole 
can be written as
\ba
T_{\pi\Sgs\rw\pi\Sgs}&=&\frac{g_{\pi\Sgs}^2}{z-z_R}\nonumber\\
T_{\bk N\rw\pi\Sgs}&=&\frac{g_{\pi\Sgs}\ g_{\bk N}}{z-z_R}\nonumber\\
T_{\pi\Sg\rw\pi\Sgs}&=&\frac{g_{\pi\Sgs}\ g_{\pi\Sg}}{z-z_R}
\ea
where the couplings $g_{\pi\Sgs}$, $g_{\bk N}$ and $g_{\pi\Sg}$ 
can be obtained from the residues at the pole. 

Writing the amplitudes for the $\Ld(1520)$ decay to $\bk N$ and $\pi\Sg$
respectively as,
\ba
-it_{\Ld(1520)\rw\bk N}=-ig_{\bk N}\ {\cal C}(1/2\ 2\
3/2;m,M-m)Y^*_{2,m-M}(\hat{k})(-1)^{M-m}\sqrt{4\pi}\nonumber\\
-it_{\Ld(1520)\rw\pi\Sg}=-ig_{\pi\Sg}\ {\cal C}(1/2\ 2\
3/2;m,M-m)Y^*_{2,m-M}(\hat{k})(-1)^{M-m}\sqrt{4\pi}
\ea
the partial decay widths of the $\Ld(1520)$ are obtained as,
\ba
\Gamma_{\bk N}&=&\frac{g_{\bk N}^2}{2\pi}\ \frac{M_N}{M_{\Ld}}\ k_{\bk}\nonumber\\
\Gamma_{\pi\Sg}&=&\frac{g_{\pi\Sg}^2}{2\pi}\ \frac{M_{\Sg}}{M_{\Ld}}\ k_\pi
\ea
where $k_{\bk}=|\vec{q}_{on}|=242$ MeV and $k_\pi=|\vec{q}_{on}\pr|=263$ MeV.
The partial decay
width to the $\pi\Sgs$ channel is zero because the $\Ld(1520)$ pole is below the threshold
for this channel. Note that $g_{\bk N}$
and $g_{\pi\Sg}$ automatically incorporate the $\beta_{\bk N}|\vec{q}_{on}|^2$
and $\beta_{\pi\Sg}|\vec{q}_{on}\pr|^2$ of the transition potential since at
least one $\pi\Sgs\rw\bk N$ transition is needed in the Bethe Salpeter series.
Hence the term $|\vec{q}_{on}|^4\ k_{\bk}=k_{\bk}^5$ guarantees the $d$-wave character
of the decay. 

\begin{figure}[h]
\includegraphics[width=0.5\textwidth]{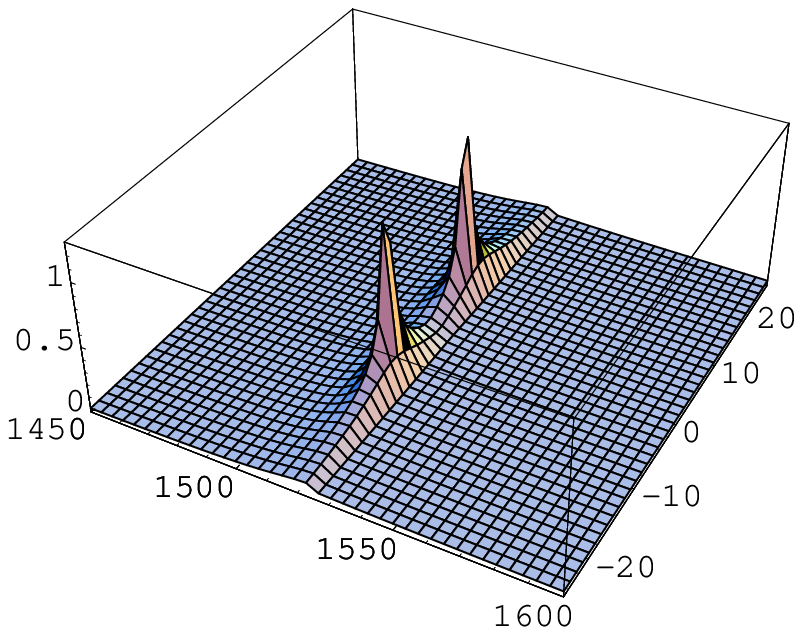}
\includegraphics[width=0.5\textwidth]{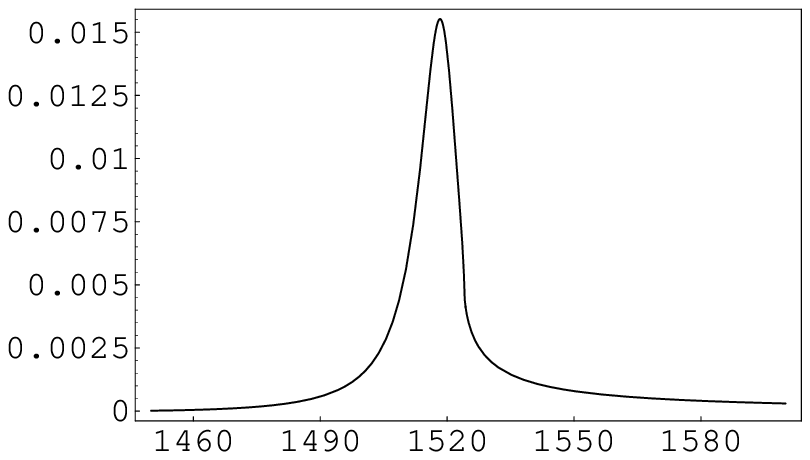}
\caption{(Color online) Left: The $\Ld(1520)$ pole as seen in the $\bk N\rw\pi\Sgs$ 
amplitude in the complex $\sqrt{s}$ plane. Right:$|T_{\bk N\rw\pi\Sgs}|^2$ in
(MeV$^{-2}$).}
\label{polefig2}
\end{figure}

We vary $\beta_{\bk N}$ and $\beta_{\pi\Sg}$ 
to reproduce the correct partial decay widths of the $\Ld(1520)$ into $\bk
N(45\%)
$ and $\pi\Sg(42\%)
$ out of a total width of 15.6 MeV and simultaneously the subtraction constant
$a$ in order to have the pole at the experimental $\Ld(1520)$ mass. This exercise results in the values
\(|g_{\pi\Sgs}|=1.57, \ |g_{\bk N}|=0.54\) and \(|g_{\pi\Sg}|=0.45\) for the couplings
of the various channels to $\Ld(1520)$ using 
$\beta_{\bk N}=2.4\times 10^{-7}$, $\beta_{\pi\Sg}=1.7\times 10^{-7}$ in units of
MeV$^{-3}$ and $a=-2.5$, fixing $\mu=700$ MeV. With this we obtain the 
$\Ld(1520)$ pole at the position $z_R=1519.7-i7.9$ as seen in
fig.~\ref{polefig2}.  

The isoscalar part of the amplitudes for specific charge channels can be 
obtained using\footnote{we use $|K^-\rgl=-|\frac{1}{2}\,
\frac{1}{2}\rgl$ and $|\Sg^+\rgl=-|1\,1\rgl$}
\ba
|\pi\Sg ;\ I=0\rgl&=&-\frac{1}{\sqrt{3}}\ |\pi^-\Sigma^{+}\rgl
-\frac{1}{\sqrt{3}}\ |\pi^0\Sigma^{0}\rgl
-\frac{1}{\sqrt{3}}\ |\pi^+\Sigma^{-}\rgl\nonumber\\
|\bk N ;\ I=0\rgl&=&\frac{1}{\sqrt{2}}\ |\bk^0 n\rgl
+\frac{1}{\sqrt{2}}\ |K^-p\rgl
\ea
and multiplying the $I=0$ amplitudes obtained above by the relevant CG coefficients.
It is to be noted that $\beta_{\bk N}$ and $\beta_{\pi\Sg}$ have been fitted to the partial
decay widths. Hence we are not making any prediction for these couplings, or
equivalently $g_{\bk N}$ and $g_{\pi\Sg}$. However, the
coupling $g_{\pi\Sgs}$ is a prediction of the theory, up to small changes
in the fine tuning of the subtraction constant.

\section{The reaction $K^-p\rw\pi^0\Sigma^{*0}(1385)\rw\pi^0\pi^0\Ld(1116)$}

\begin{figure}[h]
\centerline{\includegraphics[width=0.5\textwidth]{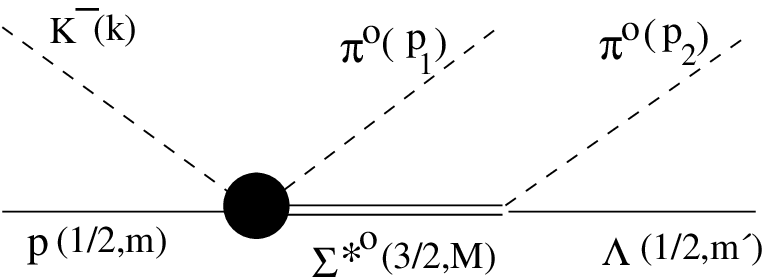}}
\caption{Scheme for $K^-p\rw\pi^0\Sigma^{*0}(1385)\rw\pi^0\pi^0\Ld(1116)$. The
blob indicates the unitarized vertex.}
\label{kpfig}
\end{figure}

Here we evaluate the cross-section for the reaction
$K^-p\rw\pi^0\Sigma^{*0}$ generated by the coupled channel scheme and the
subsequent decay of the $\Sigma^{*0}(1385)$ to $\pi^0\Ld(1116)$  
as shown in fig.~\ref{kpfig}.

To obtain the cross-section for $K^-p\rw\pi^0\Sigma^{*0}$ in the $K^-p$ CM 
frame we use the formula
\be
\frac{d\sigma}{d\Omega}=\frac{1}{16\pi^2}\ \frac{M_NM_{\Sgs}}{s}\  
\frac{|\vec p_1|}{k}\ \overline{\sum_i}\sum_f \ |t_{K^- p\rw \pi^0 \Sigma^{*0}}|^2
\ee
where $|\vec p_1|$ and $\vec
k=(0,0,k)$ denote the momenta of the outgoing pion and the incoming kaon
respectively. 
Using Eq.~(\ref{actual}) and \(Y_{2,m-M}(\hat k)=\sqrt{\frac{5}{4\pi}}\
\delta_{mM}\) and taking into account the CG coefficients we find
\be
t_{K^- p\rw \pi^0 \Sigma^{*0}}=\sqrt{\frac{1}{3}}\ T_{\bk N\rw\pi\Sgs}\ 
\delta_{mM}\ \left\{\begin{array}{l}{-1~~~~~~m=+1/2}\\{+1~~~~~~m=-1/2}
\end{array}\right\}
\ee
where $m$ is the spin of the proton and $M$ that of the $\Sg^{*0}$.
The cross section is then given by
\be
\sigma=\frac{1}{12\pi}\ \frac{M_NM_{\Sgs}}{s}\ \frac{|\vec p_1|}{k}\
|T_{\bk N\rw\pi\Sgs}|^2~.
\ee

To obtain the cross section for $K^-p\rw\pi^0\Sigma^{*0}\rw\pi^0\pi^0\Ld$ 
we now evaluate the Feynman diagram of fig.~\ref{kpfig}
where the $\Sg^{*0}$ appears as a particle propagator.

The vertex $\Sg^{*0}\rw\pi^0\Ld$ is given by~\cite{angels_phi}
\be
-it_{\pi^0\Ld\rw\Sg^{*0}}=-\frac{f_{\Sgs\pi\Ld}}{m_\pi}\ \vec S^\dagger\cdot
\vec p_2\pr
\ee 
where $S^\dagger$ is the 1/2 to 3/2 spin transition operator
and the coupling $f_{\Sgs\pi\Ld}$ is fitted to the partial decay width
of 32 MeV for $\Sigma^{*0}\rw\pi^0\Ld$. Using the $SU(3)$ arguments
of~\cite{angels_phi} one obtains $\frac{f_{\Sgs\pi\Ld}}{m_\pi}
=\frac{6}{5}\frac{D+F}{2f}$.
The amplitude for
the process shown in fig.~\ref{kpfig} in the $K^-p$ CM is then obtained as
\be
-it(\vec p_1,\vec p_2)=\frac{-iT_{\bk N\rw\pi\Sgs}}{3\sqrt{2}}\ 
\frac{f_{\Sgs\pi\Ld}/m_\pi}{M_R-M_{\Sgs}+i\Gamma_{\Sgs}(M_R)/2}\ 
\left\{\begin{array}{l}{-2p_{2z}\pr~~~~~~~~~m\pr=+1/2}\\
{p_{2x}\pr+ip_{2y}\pr~~~~m\pr=-1/2}
\end{array}\right\} 
\ee
where $m\pr$ is the spin of the outgoing $\Ld$.
Here, and in the following we will take
the spin projection $m=+1/2$ for the proton.
The $p$-wave decay width of the propagating $\Sgs$ is 
given by
\be
\Gamma_{\Sgs}(M_R)=\frac{1}{6\pi}\frac{f_{\Sgs\pi\Ld}^2}{m_\pi^2}\ \frac{M_\Ld}{M_R}\ 
|\vec p_2\pr|^3
\ee 
from where we obtain $f_{\Sgs\pi\Ld}=1.3$ for $M_R=M_{\Sgs}$,
which only differs from the $SU(3)$ value given above
by about $10\%
$.
The momentum $\vec p_2\pr$ of the final pion in the rest system of 
the $\Sgs$ is obtained as
\be
\vec p_2\pr=\left[\left(\frac{E_R}{M_R}-1\right)\ \frac{(\vec p_2\cdot\vec p_1)}
{|\vec p_1|^2}+\frac{p_2^0}{M_R}\right]\vec p_1
+\vec p_2
\label{boost}
\ee
where $M_R^2=(p_2+p_\Ld)^2$ and $E_R^2=\vec p_1^2+M_R^2$.

\begin{figure}[h]
\centerline{\includegraphics[width=0.6\textwidth]{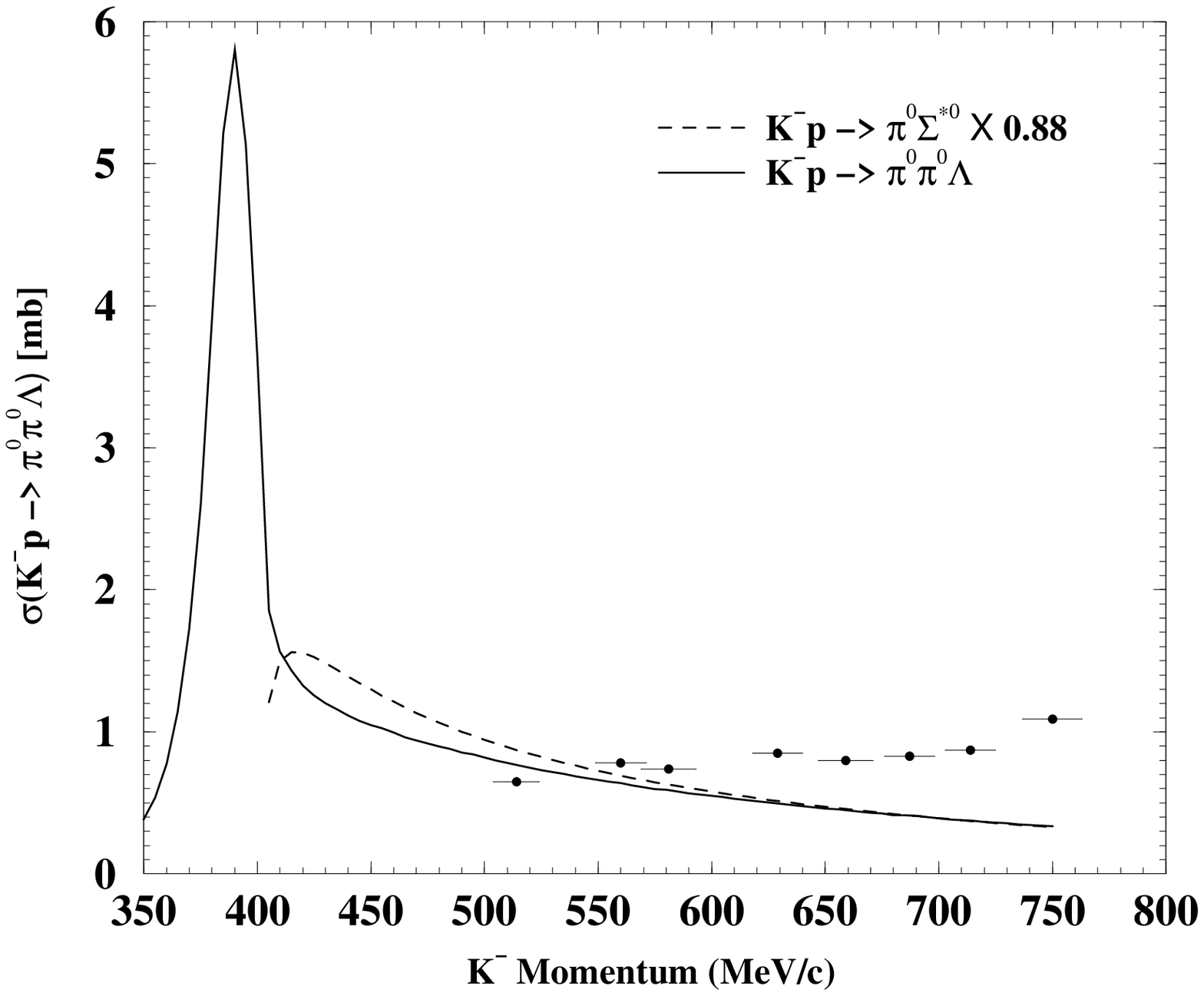}}
\caption{Cross-section as a function of the $K^-$ momentum}
\label{sigfig1}
\end{figure}

In the next step, the total squared amplitude for $K^-p\rw\pi^0\pi^0\Ld$ is
 symmetrized 
in the momenta $\vec p_1$ and $\vec p_2$ to account for the two
$\pi^0$s in the final state so that,
\be
|Amp|^2=\sum_{m\pr}|t(\vec p_1,\vec p_2)+t(\vec p_2,\vec p_1)|^2~.
\ee
The cross section is then obtained by integrating the above amplitude 
over the three-particle phase space (with a factor
1/2 for the identity of the two pions). Details are discussed in the appendix.
The results are shown in fig.~\ref{sigfig1}. The peak in the cross section for
$K^-p\rw\pi^0\pi^0\Ld$ (solid line) corresponds to the $\Ld(1520)$. We observe
a fair agreement with the experimental data~\cite{data} in the region of $K^-$ 
momenta
up to about 600 MeV from where other mechanisms for $\pi^0\pi^0\Ld$ not tied to
$\pi^0\Sg^{*0}$ production become more relevant as we shall see. 
The cross section for $K^-p\rw\pi^0\Sigma^{*0}$ multiplied by 
the $\Sgs\rw\Ld\pi$ branching ratio (=0.88) 
is also shown for comparison (dashed line). Recall that the threshold for this reaction lies
just above the peak of the $\Ld(1520)$. It is interesting to see that there is a
good agreement between the two methods of calculation when we are above the
$\pi^0\Sg^{*0}$ threshold. However, evaluating the $K^-p\rw\pi^0\Sg^{*0}$
cross section, assuming the $\Sg^{*0}$ as a stable particle gives no cross
section below the $\pi^0\Sg^{*0}$ threshold and then the explicit evaluation of
$K^-p\rw\pi^0\pi^0\Ld$ using the $\Sg^{*0}$ propagator becomes mandatory and
 provides strength below
this threshold. This feature is rather interesting because one can see the shape
of the $\Ld(1520)$ in the cross section as a function of the $K^-$ momentum. 
The strength of this peak is a genuine prediction of the theory, as well as 
the strength predicted around
500--600 MeV/c $K^-$ momentum. 

It would be instructive to get
data around the energy of the peak, since it would be a clean proof of the
link between the $\Ld(1520)$ and the $\pi\Sgs$ channel which is the basic
prediction of the chiral unitary approach.

\begin{figure}[h]
\centerline{\includegraphics[width=0.8\textwidth]{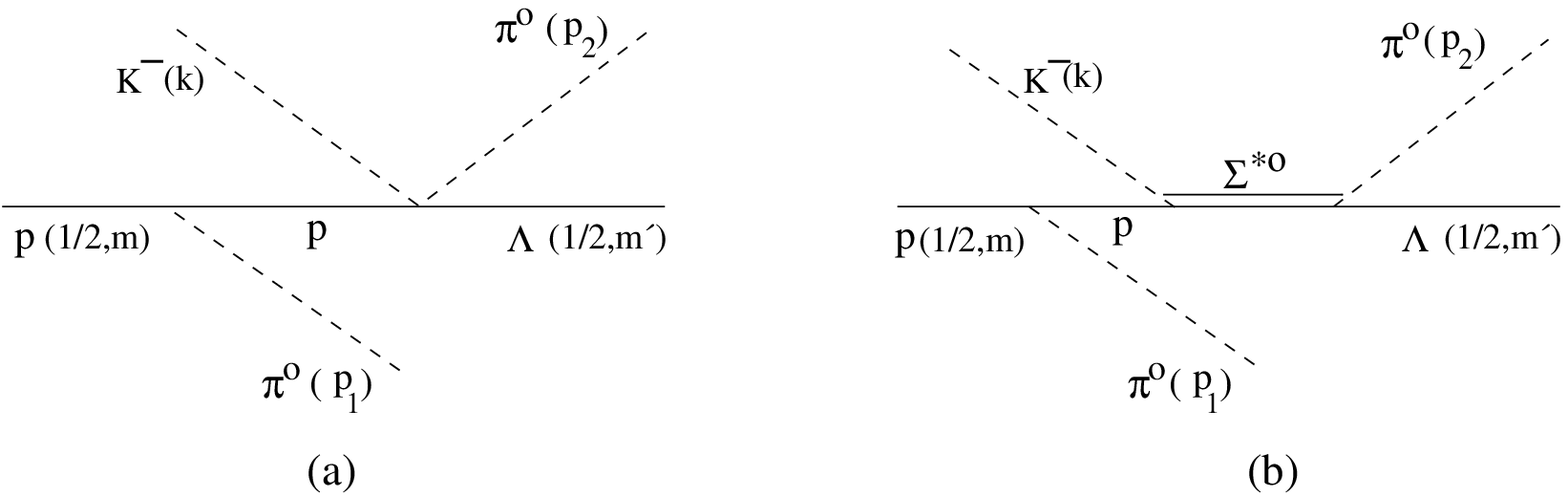}}
\caption{A conventional scheme for $K^-p\rw\pi^0\pi^0\Ld$}
\label{conv1}
\end{figure}

\begin{figure}[h]
\centerline{\includegraphics[width=.8\textwidth]{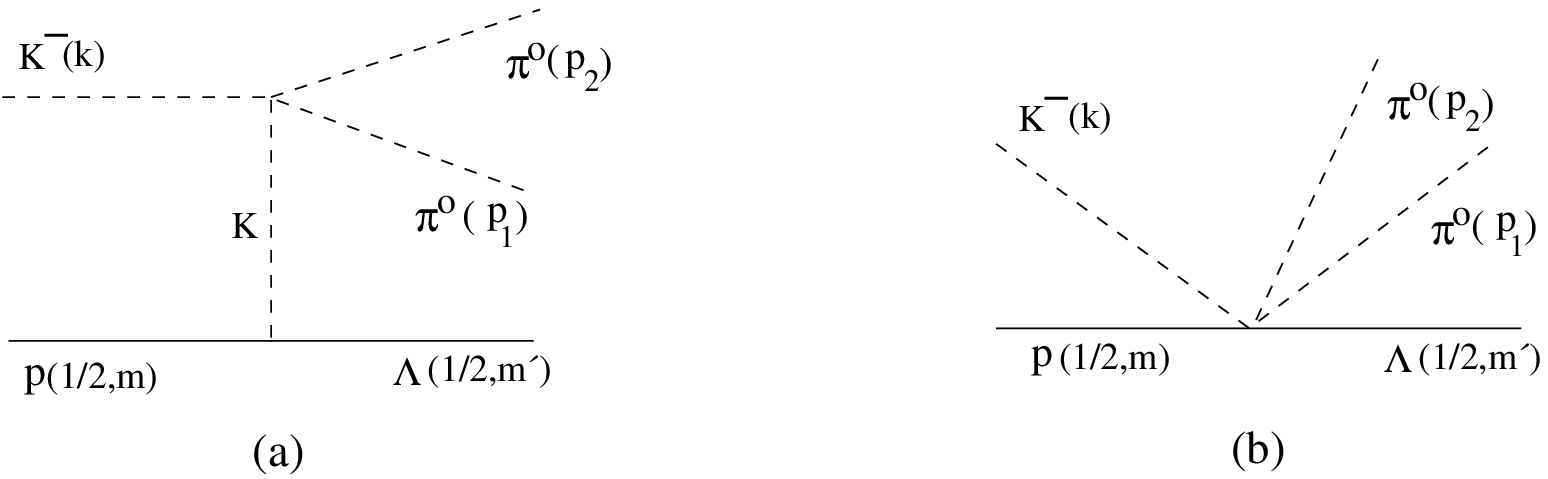}}
\caption{A conventional scheme for $K^-p\rw\pi^0\pi^0\Ld$}
\label{conv2}
\end{figure}

We will now consider other mechanisms, figs.~\ref{conv1} and \ref{conv2} which
are not tied to the $\Ld(1520)$ resonance. In
fig.~\ref{conv1} we separate the $K^-p\rw\pi^0\Ld$ interaction in $s$-wave (a)
and $p$-wave (b), this latter one dominated by the $\Sgs$
pole~\cite{jido_pwave}.
Since there is no $s$-wave resonance in $\pi^0\Ld$ around the energies we investigate, it
is enough to take for $K^-p\rw\pi^0\Ld$ the lowest order chiral amplitude
in $s$-wave in fig.~\ref{conv1}(a), which we get from~\cite{angels}, and we obtain for the
amplitude of this diagram,
\be
-it^{(s-wave)}=\frac{\sqrt{3}}{2}\ \frac{1}{4f^2}\frac{D+F}{2f}
\frac{k^{0\,\prime}+p_2^{0\,\prime}}{E_N(\vec k)-p_1^0-E_N(\vec k+\vec p_1)}
\left\{\begin{array}{l}{p_{1z}~~~~~~~~~~~~~m\pr=+1/2}\\{p_{1x}+ip_{1y}~~~~m\pr=-1/2}
\end{array}\right\}
\label{swave}
\ee
where $k^{0\,\prime}$ and $p_2^{0\,\prime}$ are the energies of $\vec k$ and
$\vec p_2$ written in the $\pi^0\Ld$ CM frame.

The amplitude 
corresponding to the diagram of fig.~\ref{conv1}(b), is given by
\ba
-it^{(p-wave)}&=&-\frac{D+F}{2f}\ \frac{f_{\Sgs\pi\Ld}}{m_\pi}\ 
\frac{f_{K^-p\Sg^{*0}}}{m_\pi}\ \ 
\vec S^\dagger\cdot\vec p_2\pr\ \vec S\cdot\vec k\pr\ \vec\sigma\cdot\vec
p_1\nonumber\\
&&\times\frac{1}{M_R-M_{\Sgs}+i\Gamma_{\Sgs}(M_R)/2}\ 
\frac{1}{E_N(\vec k)-p_1^0-E_N(\vec k+\vec p_1)}
\ea
where $f_{K^-p\Sg^{*0}}$ is given in~\cite{angels_phi} by
\be
\frac{f_{K^-p\Sg^{*0}}}{m_\pi}=-\frac{2\sqrt{3}}{5}\frac{D+F}{2f}.
\ee
and
\begin{eqnarray*}
&&
\vec S^\dagger\cdot\vec p_2\pr\ \vec S\cdot\vec k\pr\ \vec\sigma\cdot\vec
p_1=\nonumber\\
&&\left\{\begin{array}{l}
-\frac{i}{3}\, (\vec p_2\pr\times \vec k\pr)\,\cdot\vec p_1
+\frac{2}{3}\, (\vec p_2\pr\cdot\vec k\pr)\, p_{1z}
+\frac{1}{3}\, (\vec p_1\cdot\vec p_2\pr)\, k_z\pr
-\frac{1}{3}\, (\vec p_1\cdot\vec k\pr)\, p_{2z}\pr~~~~~~~~~~~m\pr=+1/2\\
\frac{2}{3}\, (\vec p_2\pr\cdot\vec k\pr)\, (p_{1x}+ip_{1y})
+\frac{1}{3}\, (\vec p_1\cdot\vec p_2\pr)\, (k_{x}\pr+ik_{y}\pr)
-\frac{1}{3}\, (\vec p_1\cdot\vec k\pr)\, (p_{2x}\pr+ip_{2y}\pr)~~~~m\pr=-1/2
\end{array}\right\}
\end{eqnarray*}
where the boosted momenta $\vec p_2\pr$ and $\vec k\pr$ are obtained as in
Eq.~(\ref{boost}).
Next we study the amplitude corresponding to fig.~\ref{conv2}. As shown in~\cite{hyodo}
the contact term of fig.~\ref{conv2}(b) just cancels the part of 
fig.~\ref{conv2}(a) which comes from the off shell part of the meson meson
amplitude. Hence, using the diagram of fig.~\ref{conv2}(a) with the meson meson
amplitude calculated on shell accounts for the sum of the two diagrams. We
take the $K^-K^+\rw\pi^0\pi^0$ amplitude from~\cite{ollernpa97} and the $K^-p\Ld$
vertex from~\cite{angels} and we obtain
\ba
-it^{(K-pole)}&=&-\frac{1}{4f^2}\left(-\frac{2}
{\sqrt{3}}\frac{D+F}{2f}+\frac{1}{\sqrt{3}}\frac{D-F}{2f}\right)
\frac{(p_1+p_2)^2}{(k-p_1-p_2)^2-m_K^2}\nonumber\\
&&\times\left\{\begin{array}{l}{(k-p_{1z}-p_{2z})~~~~~~~~~~~~~~~~~~~~m\pr=+1/2}
\\{-(p_{1x}+p_{2x})-i(p_{1y}+p_{2y})~~~~m\pr=-1/2}
\end{array}\right\}~.
\ea

\begin{figure}[h]
\centerline{\includegraphics[width=0.6\textwidth]{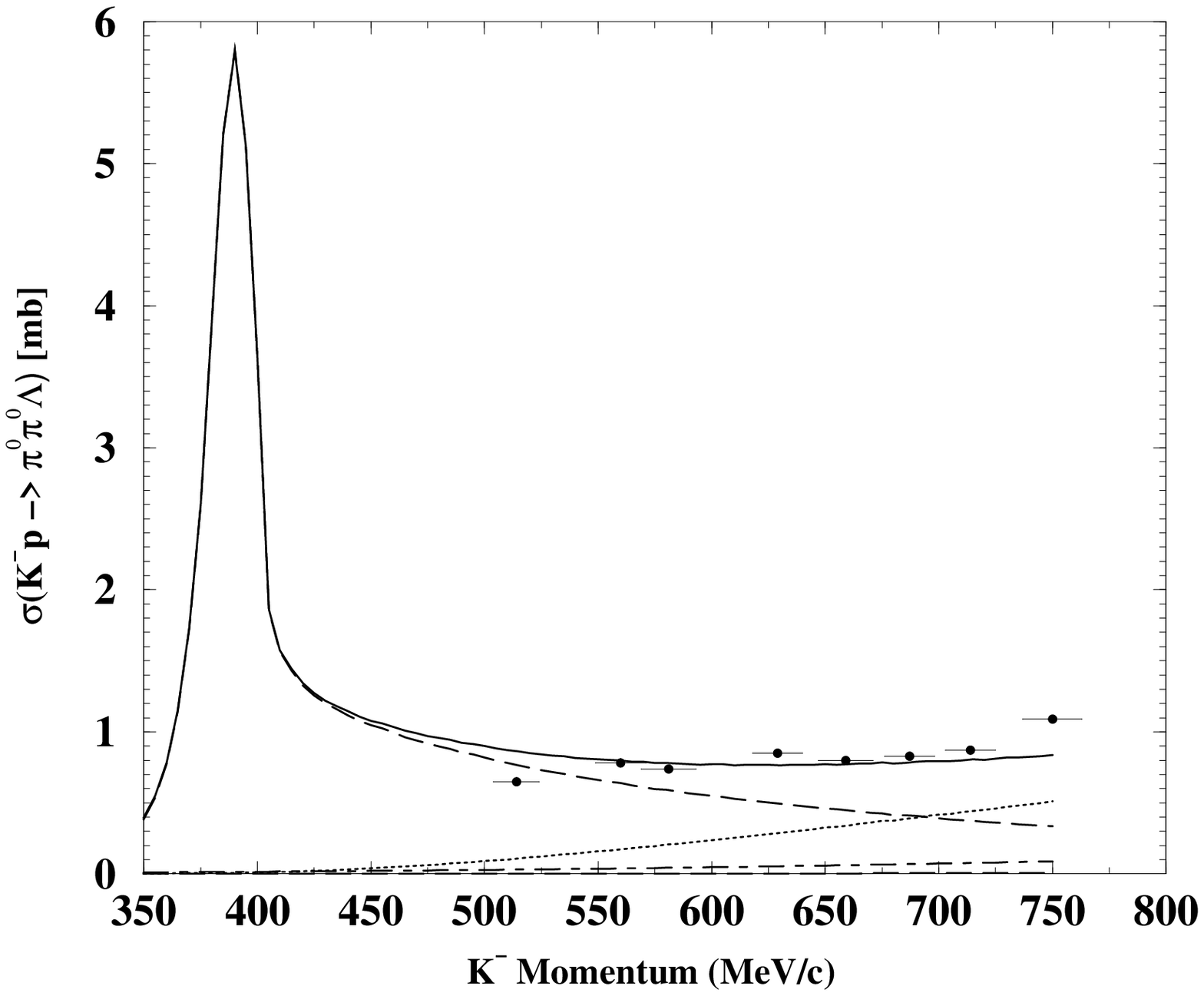}}
\caption{Cross-section as a function of the $K^-$ momentum. The dot-dashed 
and dotted lines are
the contributions of the diagrams of figs.~\ref{conv1}(a) and \ref{conv1}(b)
respectively. The
dashed line shows the cross section with fig.~\ref{kpfig} only 
and the solid line for a coherent sum of all these diagrams.}
\label{sigfig2}
\end{figure}
 
We add all these amplitudes symmetrized to the former ones and recalculate the
cross section. Note that the amplitude $t^{(K-pole)}$ is already symmetric with
respect to the momenta $p_1$ and $p_2$ and does not have to be symmetrized
again. The results are shown in
fig.~\ref{sigfig2}. We find that by themselves the new
mechanisms would give a cross section more than one order of magnitude smaller
than the experiment, up to 600 MeV/c, indicating that the dominant mechanism by far is the one
that we have investigated with the $\pi\Sgs$ tied to the $\Ld(1520)$ resonance.
Added coherently to the dominant mechanism, these new processes
produce a negligible effect around the $\Ld(1520)$ peak and they become more
visible far away from the resonance where they increase the cross section and
help to get a good agreement with the data. 

Details on the new mechanisms are as follows:

a) The kaon pole term of fig.~\ref{conv2}(a) produces a negligible effect in the
cross section not visible in fig.~\ref{sigfig2}. The $K$ propagator reduces the
strength of the diagram and the factor $(p_1+p_2)^2$ from the
$K^+K^-\rw\pi^0\pi^0$ amplitude also contributes to the small size of the term.

b) The term from the diagram of fig.~\ref{conv1}(a) involving the $s$-wave
$K^-p\rw\pi^0\Ld$ amplitude contributes about one fifth of the total cross
section at the highest energy of fig.~\ref{sigfig2} and adds practically
incoherently to the $K^-p\rw\pi^0\Sg^{*0}$ mechanism.

c) The term from the diagram of fig.~\ref{conv1}(b) involving the $p$-wave
$K^-p\rw\pi^0\Ld$ amplitude contributes about one half
 of the total cross
section at the highest energy of fig.~\ref{sigfig2} and also adds almost 
incoherently to the other mechanisms.

\begin{figure}[h]
\includegraphics[width=0.5\textwidth]{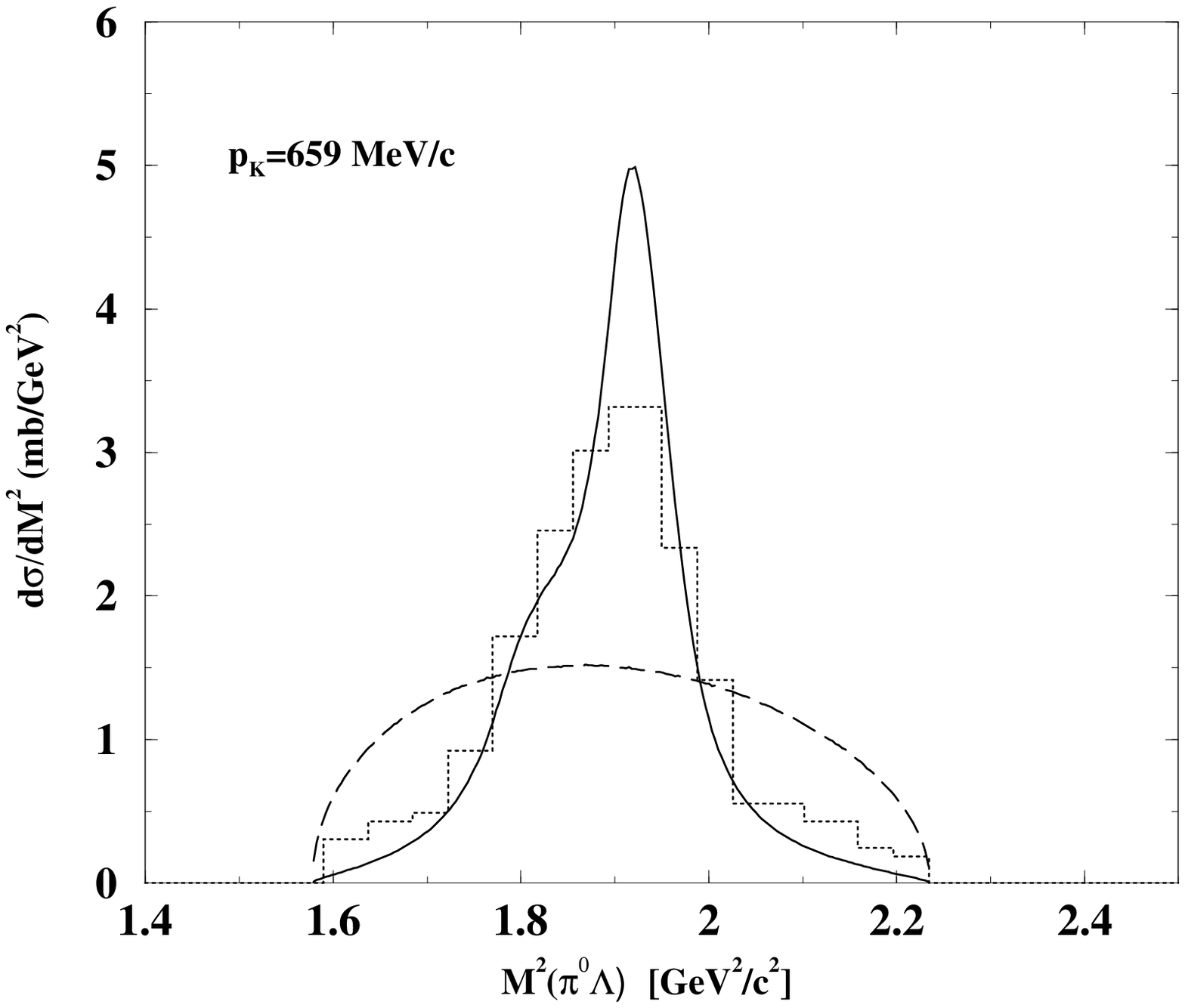}
\includegraphics[width=0.5\textwidth]{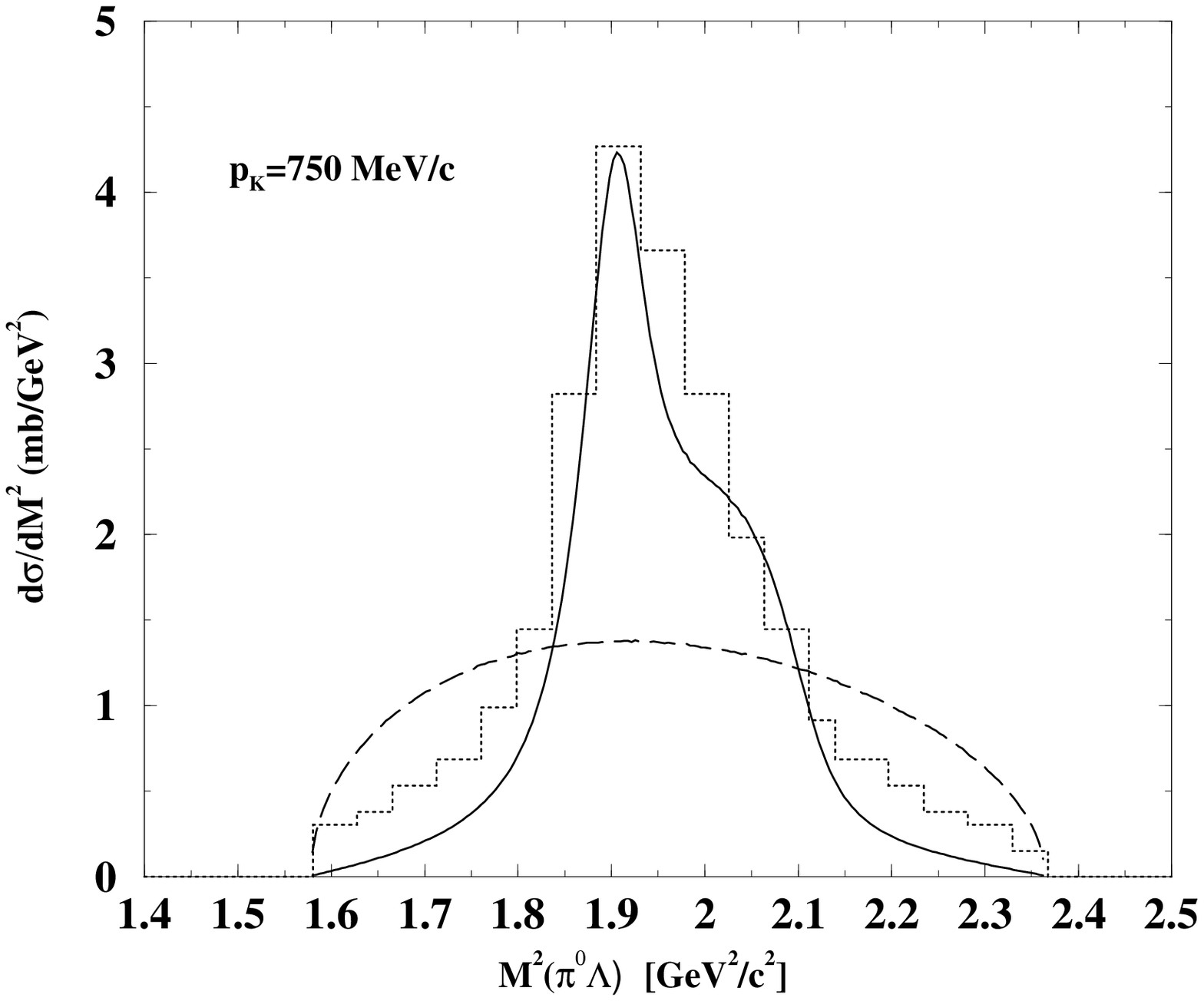}
\caption{$d\sigma/dM^2$  as a function
of the invariant mass of $\pi^0\Ld$  for two values of the $K^-$ momentum in CM;
 Left: 659 MeV
and Right:750 MeV. Solid lines represent our results. The dotted histograms are
the experimental results from~\cite{data} normalized to the total experimental
cross section. The dashed lines indicate the phase space normalized to the
theoretical cross section.}
\label{dsdmfig}
\end{figure}

We also calculate the differential cross section $d\sigma/dM^2$ as a function of
the invariant mass of a pair of $\pi^0\Ld$ for two values of $K^-$ momentum
which we plot in fig.~\ref{dsdmfig}. We
find a good agreement with the experimental curves in~\cite{data}.
In the figure we can see the $\Sgs(1385)$ peak clearly. We also notice, as
in~\cite{data}, that the effect of symmetrization of the amplitudes with respect
to the two final pions is visible in the spectra. Indeed, we see that for
$p_K$=659 MeV/c some strength piles up on the left hand side of the resonance while
for $p_K$=750 MeV/c this strength is moved to higher energies and
produces a shoulder on the right hand side. These features are also clear in the
experimental data.

\section{Conclusions} 

  We have extended the chiral unitary approach for the interaction of the decuplet
of baryons with the octet of mesons, for the case of meson baryon scattering in
the region of the $\Lambda(1520)$ resonance, by including the $\bar{K}N$ and $\pi
\Sigma$ channels which couple in $d$-wave to the main $s$-wave channels 
$\pi \Sgs(1385)$ and $K \Xi^*(1533)$. The introduction of these channels allowed us
to obtain a more realistic  description of the $\Lambda(1520)$ resonance and make
predictions for reactions which evidence the nature of this resonance as a
quasibound $\pi \Sgs(1385)$ state.  We found a good example in the 
$K^- p \to \pi \Sgs(1385) (\pi^0 \Lambda)$ reaction which has been measured
recently.  We found that the strength of the cross section was well reproduced
in terms of the large coupling of the  $\Lambda(1520)$ to $\pi \Sgs(1385)$,
which is a prediction of the chiral unitary approach. Both the total cross
sections as well as the invariant mass distributions of  $\pi^0 \Lambda$ were
well reproduced.  In addition the theory makes predictions for a large peak of
the total cross section of $K^- p \to \pi^0\pi^0 \Lambda$ for $K^- p$ energies around 
the  $\Lambda(1520)$, and hence below the  $\pi \Sgs(1385)$ threshold.  The
prediction for this cross section is related to the large coupling of the 
$\Lambda(1520)$ to $\pi \Sgs(1385)$ in spite of the fact that the 
$\pi \Sgs(1385)$ is kinematically forbidden. This region falls just below 
the data measured in the reaction that we analyze.  It is then clear that a
measurement of the reactions in  this region becomes most advisable, and
confirmation of the quantitative predictions made here would support the
idea of the $\Lambda(1520)$ as a dynamically generated resonance, and by
extension for the other resonances equally generated from the interaction of the
decuplet of baryons and octet of mesons.

\section*{Acknowledgments}
This work is partly supported by DGICYT contract number BFM2003-00856,
and the E.U. EURIDICE network contract no. HPRN-CT-2002-00311.
This research is part of the EU Integrated Infrastructure Initiative
Hadron Physics Project under contract number RII3-CT-2004-506078.

\section*{Appendix}
\setcounter{section}{0}
\renewcommand{\thesection}{\arabic{section}}
\setcounter{equation}{0}
\renewcommand{\theequation}{A.\arabic{equation}}
 
Here we describe in some detail the procedure followed to perform the
integration over the three body phase space which was encountered in the
evaluation of the cross section for the reaction 
$K^-p\rw\pi^0\pi^0\Ld$ due to the various Feynman
diagrams described in this work.

For the differential cross section we follow the definition
\ba
d\sigma&=&(2\pi)^4\delta^{(4)}(p_1+p_2+p_\Ld-k-p)\ \frac{2M_N\ 2M_\Ld}{v_{rel}\ 
2\omega_K\ 2E_N }\ S\ |Amp|^2\nonumber\\ 
&&\times\frac{d^3\vec p_1}{(2\pi)^3\ 2\omega_1}\ 
\frac{d^3\vec p_2}{(2\pi)^3\ 2\omega_2}\ \frac{d^3\vec p_\Ld}{(2\pi)^3\ 2E_\Ld}
\ea
with $v_{rel}=\left[(k\cdot p)^2-m_K^2M_N^2\right]^{1/2}/(\omega_K E_N)$ and
$S(=1/2)$, the symmetry factor for the two identical $\pi^0$.
In the $K^-p$ CM system, the cross section is obtained as
\be
\sigma=\frac{M_N\ M_\Ld}{\lambda^{1/2}(s,m_K^2,M_N^2)}\int 
\frac{d^3\vec p_1}{(2\pi)^3\ 2\omega_1}\int \frac{d^3\vec p_2}{(2\pi)^3\ 2\omega_2}
\ \frac{|Amp|^2}{2E_\Ld}\ (2\pi)\ \delta(\sqrt{s}-\omega_1-\omega_2-E_\Ld)
\ee
with $\vec p_\Ld=-(\vec p_1+\vec p_2)$ as a result of the integration over
$\vec p_\Ld$ and using $2v_{rel}\omega_K E_N=\lambda^{1/2}(s,m_K^2,M_N^2)$.
In order to simplify the angular integration we now make the following
coordinate transformation. Assuming $\phi_1=0$ without loss of generality,
let us denote by $\theta_{12}$ the angle between the vectors 
$\vec p_1$ and $\vec p_2$. We now generate the
vector $\vec p_2$ in a frame in which the polar angle is $\theta_{12}$ 
so that its components with respect to this rotated frame
are given by
\be
\vec \tp_2=\left\{\begin{array}{l}
p_2\ \sin\theta_{12}\ \cos\tph_2\\
p_2\ \sin\theta_{12}\ \sin\tph_2\\
p_2\ \cos\theta_{12}
\end{array}\right.
\ee
and the differential is given by $d^3\vec\tp_2=-|\vec\tp_2|^2\ d|\vec\tp_2|\ 
d(\cos\theta_{12})\ d\tph_2$. 
The $\delta$ function can now be used to perform the integral over 
$\cos\theta_{12}$
and we have
\be
\sigma=\frac{M_N\ M_\Ld}{\lambda^{1/2}(s,m_K^2,M_N^2)}
\ \frac{1}{8}\ \frac{1}{(2\pi)^4}
\int_{-1}^{1}d\cos\theta_1
 \int_{m_\pi}^{\omega_{max}}d\omega_1\ 
 \int_0^{2\pi}d\tph_2\ 
 \int_{m_\pi}^{\omega_{max}}d\omega_2\ 
|Amp|^2\ \Theta(1-|A|^2)
\ee
where
\[
A=\cos\theta_{12}=\frac{\left[(\sqrt{s}-\omega_1-\omega_2)^2
-|\vec p_1|^2-|\vec p_2|^2-M_\Ld^2\right]}{2\ |\vec p_1||\vec p_2|}
\]
and 
\[\omega_{max}=\frac{s+m_\pi^2-(m_\pi+M_\Ld)^2}{2\sqrt{s}}\]
is the maximum energy of the pion which is reached in the case 
when the other pion and the $\Lambda$ move together.  
The original vector $\vec p_2$ is recovered
through
\be
\vec p_2=R_y(\theta_1)\ \vec \tp_2
\ee
where 
\[
R_y(\theta_1)=\left(\begin{array}{ccc} \cos\theta_1 & 0 & \sin\theta_1\\
                                            0     & 1 &   0\\
				    -\sin\theta_1 & 0 & \cos\theta_1
		    \end{array}\right)		    
\]
is the usual rotation matrix for a rotation by an angle $\theta_1$ around the
$y$-axis. Note that $|\vec p_2|=|\vec\tp_2|$.

\begin{thebibliography}{99}

\bibitem{Eidelman:2004wy}
  S.~Eidelman {\it et al.}  [Particle Data Group],
  Phys.\ Lett.\ B {\bf 592}, 1 (2004).

\bibitem{Kaiser:1995cy}
N.~Kaiser, P.~B.~Siegel and W.~Weise,
Phys.\ Lett.\ B {\bf 362} (1995) 23

\bibitem{Kaiser:1996js}
N.~Kaiser, T.~Waas and W.~Weise,
Nucl.\ Phys.\ A {\bf 612} (1997) 297
[arXiv:hep-ph/9607459].

\bibitem{angels}
E.~Oset and A.~Ramos,
Nucl. Phys. A {\bf 635} (1998) 99.
[arXiv:nucl-th/9711022].

\bibitem{Nacher:1999vg}
J.~C.~Nacher, A.~Parreno, E.~Oset, A.~Ramos, A.~Hosaka and M.~Oka,
Nucl.\ Phys.\ A {\bf 678} (2000) 187
[arXiv:nucl-th/9906018].

\bibitem{oller}
J.~A.~Oller and U.-G.~Mei{\ss}ner,
Phys.\ Lett.\ B {\bf 500} (2001) 263.
[arXiv:hep-ph/0011146].

\bibitem{Inoue:2001ip}
T.~Inoue, E.~Oset and M.~J.~Vicente Vacas,
Phys.\ Rev.\ C {\bf 65} (2002) 035204
[arXiv:hep-ph/0110333].

\bibitem{bennhold}
 E.~Oset, A.~Ramos and C.~Bennhold,
 Phys.\ Lett.\ B {\bf 527} (2002) 99
 [Erratum-ibid.\ B {\bf 530} (2002) 260]
 [arXiv:nucl-th/0109006].
 
\bibitem{Garcia-Recio:2002td}
  C.~Garcia-Recio, J.~Nieves, E.~Ruiz Arriola and M.~J.~Vicente Vacas,
  Phys.\ Rev.\ D {\bf 67} (2003) 076009
  [arXiv:hep-ph/0210311].
  
\bibitem{Jido:2003cb}
D.~Jido, J.~A.~Oller, E.~Oset, A.~Ramos and U.~G.~Meissner,
Nucl.\ Phys.\ A {\bf 725} (2003) 181
[arXiv:nucl-th/0303062].

\bibitem{Garcia-Recio:2003ks}
C.~Garcia-Recio, M.~F.~M.~Lutz and J.~Nieves,
Phys.\ Lett.\ B {\bf 582} (2004) 49
[arXiv:nucl-th/0305100].

\bibitem{lutz}
E.~E.~Kolomeitsev and M.~F.~M.~Lutz,
Phys.\ Lett.\ B {\bf 585} (2004) 243
[arXiv:nucl-th/0305101].

\bibitem{decu_ss}
S.~Sarkar, E.~Oset and M.~J.~Vicente Vacas,
Nucl.\ Phys.\ A {\bf 750} (2005) 294
[arXiv:nucl-th/0407025].

\bibitem{review}
J.~A.~Oller, E.~Oset and A.~Ramos,
Prog.\ Part.\ Nucl.\ Phys.\  {\bf 45} (2000) 157
[arXiv:hep-ph/0002193].

\bibitem{manohar}
E.~Jenkins and A.~V.~Manohar,
Phys.\ Lett.\ B {\bf 259} (1991) 353.

\bibitem{Gasser:1984gg}
J.~Gasser and H.~Leutwyler,
Nucl.\ Phys.\ B {\bf 250} (1985) 465.

\bibitem{savage} 
M.~N.~Butler, M.~J.~Savage and R.~P.~Springer,
Nucl.\ Phys.\ B {\bf 399}, 69 (1993)
[arXiv:hep-ph/9211247].

\bibitem{nsd}
 J.~A.~Oller and E.~Oset,
 Phys.\ Rev.\ D {\bf 60} (1999) 074023
 [arXiv:hep-ph/9809337].

\bibitem{angels_phi}
 E.~Oset and A.~Ramos,
 Nucl.\ Phys.\ A {\bf 679} (2001) 616
 [arXiv:nucl-th/0005046].

\bibitem{data}
S.~Prakhov {\it et al.},
Phys.\ Rev.\ C {\bf 69} (2004) 042202.

\bibitem{jido_pwave}
 D.~Jido, E.~Oset and A.~Ramos,
 Phys.\ Rev.\ C {\bf 66} (2002) 055203
 [arXiv:nucl-th/0208010].
 
\bibitem{hyodo}
T.~Hyodo, A.~Hosaka, E.~Oset, A.~Ramos and M.~J.~Vicente Vacas,
Phys.\ Rev.\ C {\bf 68} (2003) 065203
[arXiv:nucl-th/0307005].

\bibitem{ollernpa97}
J.~A.~Oller and E.~Oset,
Nucl.\ Phys.\ A {\bf 620} (1997) 438
[Erratum-ibid.\ A {\bf 652} (1999) 407]
[arXiv:hep-ph/9702314].

\end{thebibliography}
\end{document}